\documentclass[twocolumn]{aa}
\usepackage{astrobib}
\usepackage{journals}
\usepackage{graphicx}
\usepackage{txfonts}
\begin{document}
   \title{The Proper Motion of the Local Group Galaxy IC\,10}

   \subtitle{}

   \author{A. Brunthaler
          \inst{1}
          \and
          M.J. Reid\inst{2}  
          \and 
          H. Falcke\inst{3,4}
          \and
          C. Henkel\inst{1}
          \and
          K.M. Menten\inst{1}
          }

   \offprints{brunthal@mpifr-bonn.mpg.de}

   \institute{Max-Planck-Institut f\"ur Radioastronomie, Auf dem H\"ugel 69,
              53121 Bonn, Germany
              \and
              Harvard-Smithsonian Center for Astrophysics, 60 Garden Street,
              Cambridge, MA 02138, USA
              \and
              ASTRON, Postbus 2, 7990 AA Dwingeloo, the Netherlands
              \and 
               Department of Astrophysics, Radboud Universiteit
               Nijmegen, Postbus 9010, 6500 GL Nijmegen, The Netherlands}

   \date{Received; accepted}

   \abstract{We have measured the proper motion of the Local Group galaxy 
             IC\,10 with the Very Long Baseline Array by measuring the position
             of an H$_2$O maser relative to two background quasars over 4.3 
             years. After correction for the rotation of the Milky Way and 
             IC\,10, we obtain a motion of --39$\pm$9 $\mu$as~yr$^{-1}$ toward
             the East and 31$\pm$8 $\mu$as~yr$^{-1}$  toward the North. 
             This corresponds to a total space velocity of 215 $\pm$ 42 
             km~s$^{-1}$ relative to the Milky Way for an assumed distance 
             of 660$\pm$66 kpc. Assuming that IC\,10 and M33, for which also a
             proper motion measurement exists, are bound to M31, we calculate 
             a lower limit for the mass of M31 of 
             7.5~$\times 10^{11}$M$_\odot$.

   }

   \maketitle
%

\section{Introduction}
\subsection{Proper Motions in the Local Group}

Proper motion measurements of Local Group galaxies are important for our 
understanding of the dynamics and evolution of the Local Group. Presently, 
measurements of extragalactic proper motions by optical telescopes are 
limited to the most nearby companions of the Milky Way, i.e. the LMC 
\cite{JonesKlemoaLin1994,KallivayalilvanderMarelAlcock2006a,PedrerosCostaMendez2006}, the SMC \cite{KallivayalilvanderMarelAlcock2006b}, 
the Sculptor dwarf spheroidal galaxy (dShp) 
\cite{SchweitzerCurworthMajewski1995,PiatekPryorBristow2006}, the Canis Major 
dwarf galaxy \cite{DinescuMartinez-DelgadoGirard2005}, the Ursa Minor~dSph 
\cite{PiatekPryorBristow2005}, the Sagittarius~dSph 
\cite{DinescuGirardvanAltena2005}, the Fornax~dSph 
\cite{PiatekPryorOlszewski2002,DinescuKeeneyMajewski2004}, and the Carina~dSph 
\cite{PiatekPryorOlszewski2003}. These galaxies are all closer than 150 kpc 
and show motions between 0.2 and a few milliarcseconds (mas) per year. More 
distant galaxies, such as galaxies in the Andromeda subgroup at distances of 
$\sim$ 800 kpc, have smaller angular motions, which are currently not 
measurable with optical telescopes. 

On the other hand, \citeN{BrunthalerReidFalcke2005} measured the 
proper motions of two groups of water masers on opposite sides of M33
at radio frequencies with the NRAO\footnote{The National Radio Astronomy 
Observatory is operated by Associated Universities, Inc., under a cooperative 
agreement with the National Science Foundation.} Very Long Baseline Array 
(VLBA). A comparison of the relative proper motion between 
the two groups of masers and their expected motions from the known rotation curve and 
inclination of M33 led to a determination of a ``rotational parallax'' 
(730 $\pm$ 168 kiloparsec) for this galaxy. This distance is consistent 
with recent Cepheid and tip of the red giant branch estimates 
(\citeNP{LeeKimSarajedini2002}; \citeNP{McConnachieIrwinFerguson2005}) and 
earlier distance estimates using the internal motions of water masers in 
the IC\,133 region \cite{GreenhillMoranReid1993,ArgonGreenhillMoran2004}. 

Since 
the proper motion measurements were made relative to a distant extragalactic 
background source, the proper motion of M33 itself could also be determined.  
This measured proper motion of M33 is a first important step toward a 
kinematical model of the Local Group and was used to constrain the proper 
motion and dark matter content of the Andromeda Galaxy M31 \cite{LoebReidBrunthaler2005}.

Water masers in  Local Group galaxies have also been found toward the 
Magellanic Clouds (e.g.~\citeNP{ScaliseBraz1981}) and IC\,10 (e.g. 
\citeNP{HenkelWouterlootBally1986}). Other Local Group galaxies were searched, 
but no additional water masers have been detected (see 
\citeNP{BrunthalerHenkeldeBlok2006} and references therein). In this paper
we report on VLBA observations of the maser in IC\,10 to measure its motion.

\subsection{IC\,10}

The extragalactic nature of IC\,10 was first recognized by
\citeN{Mayall1935}. ~\citeN{Hubble1936} proposed that it was likely a
member of the Local Group and described it as ``one of the most curious
objects in the sky''. However, observations of IC\,10 have been always
difficult because of the low Galactic latitude of 3$^\circ$. IC\,10
has been classified as an Ir\,IV galaxy (e.g. \citeNP{vandenBergh1999}),
but \citeN{RicherBullejosBorissova2001} argue that it has more
properties of a blue compact dwarf galaxy. It is also the nearest galaxy 
hosting a small starburst, evidenced by its large number of Wolf-Rayet stars
(\citeNP{MasseyArmandroffConti1992}) and the discovery of 144 H\,II
regions (\citeNP{HodgeLee1990}). Observations of H\,I with  the
Westerbork Synthesis Radio Telescope by~\citeN{ShostakSkillman1989}
revealed that IC\,10 has a regularly rotating disk surrounded by a
counter-rotating outer distribution of gas.

The distance to IC\,10 is subject to controversy because of 
difficulties caused by its low Galactic latitude. Early estimates
claim a distance of 1--1.5 Mpc \cite{Roberts1962} and 3 Mpc
(\citeNP{BottinelliGouguenheimHeidmann1972}, \citeNP{SandageTammann1974}). 
\citeN{Huchtmeier1979} argued for a closer distance of 1 Mpc. The most recent 
determination from multi-wavelength observations of Cepheid variables  
obtained a distance of 660$\pm$66 kpc to IC\,10 \cite{SakaiMadoreFreedman1999},
which we adopt throughout this paper.

IC\,10 hosts two known H$_2$O masers, IC\,10-SE and IC\,10-NW 
\cite{BeckerHenkelWilson1993}. The strong SE-component was first detected by 
~\citeN{HenkelWouterlootBally1986} and the whole spectrum of IC\,10 showed 
strong variability since its discovery with flux densities between less than
1 Jy \cite{BeckerHenkelWilson1993} and a flare with a (single dish) 
flux density of 125 Jy \cite{BaanHaschick1994}. Even intraday variability
has been reported by~\citeN{ArgonGreenhillMoran1994}, but the strong
component at $v_\mathrm{LSR}\approx -324$~km~s$^{-1}$ has been
 persistent until now.

\section{Observations and Data Reduction}

We observed the 
usually brightest maser in IC\,10-SE with the VLBA thirteen times 
between February 2001 and June 2005. The observations are grouped into six 
epochs, each comprising two closely spaced observations,
except the first epoch with three observations, to enable assessment 
of overall accuracy and systematic errors (Table~\ref{obsinfo}).

\begin{table}
      \caption[]{Details of the observations: Observing date,
      observation length $t_{obs}$, beam size $\theta$ and position
      angle $PA$.}
         \label{obsinfo}
      \[
  \begin{tabular}{p{0.08\linewidth}cp{0.10\linewidth}cp{0.10\linewidth}cp{0.10\l
inewidth}cp{0.10\linewidth}cp{0.08\linewidth}}
           \hline
Epoch& Date & $t_{obs} [h]$  & $\theta$ [mas] &$PA [^\circ]$\\
            \hline
     I&      2001/02/09 & 10 & 0.53$\times$0.33& -15 \\
     I&      2001/03/28 & 10 & 0.55$\times$0.36& -18 \\
     I&      2001/04/12 & 10 & 0.63$\times$0.37&  -5 \\
            \hline
     II&     2002/01/12 & 10 & 0.59$\times$0.35& -19 \\ 
     II&     2002/01/17 & 10 & 0.64$\times$0.32& -22 \\  
            \hline
     III&    2002/10/01 & 10 & 0.68$\times$0.38& -12 \\
     III&    2002/10/11 & 10 & 0.61$\times$0.34&  -5\\
            \hline
      IV&    2003/12/12 & 12 & 0.52$\times$0.33& -15\\ 
      IV&    2004/01/10 & 12 & 0.50$\times$0.33& -23\\ 
            \hline
       V&    2004/08/23 & 12 & 0.60$\times$0.51&  -2\\     
       V&    2004/09/18 & 12 & 0.54$\times$0.35& -17\\ 
            \hline
      VI&    2005/06/01 & 12 & 0.60$\times$0.50& -11\\
      VI&    2005/06/07 & 12 & 0.56$\times$0.39&  -6\\ 
         \end{tabular}
      \]
   \end{table}

We observed in four 8 MHz bands in dual circular polarization. The 128 spectral 
channels in each band yielded a channel
spacing of 62.5 kHz, equivalent to 0.84 km s$^{-1}$, and covered a
velocity range of 107 km s$^{-1}$.  The observations involved rapid
switching between the phase-calibrator VCS1~J0027+5958 from the VLBA Calibrator
Survey \cite{BeasleyGordonPeck2002}, which is a compact background source 
with continuum emission, and the target sources IC\,10 and NVSS~J002108+591132.
NVSS~J002108+591132 is a radio continuum source from the NRAO VLA Sky Survey 
(NVSS) \cite{CondonCottonGreisen1998} and is located only 8 arcminutes from 
the maser in IC\,10. It was also detected in X-rays 
\cite{WangWhitakerWilliams2005} and is most likely also a background quasar. 
The redshifts of VCS1~J0027+5958 and NVSS~J002108+591132 are not known. We 
switched sources every 30 seconds in the sequence VCS1~J0027+5958 -- IC\,10 
-- VCS1~J0027+5958 -- NVSS~J002108+591132 -- VCS1~J0027+5958 and achieved  
on-source integration times of $\sim$ 22 seconds. The background sources were assumed to 
be stationary on the sky. Since the phase-calibrator is separated by only 
1$^\circ$ on the sky from the target sources, one can obtain precise angular 
separation measurements.

From the second epoch on, we included {\it geodetic-like} observations where
we observed for 45 minutes 10--15 strong radio sources ($>$ 200 mJy) with 
accurate positions ($<$ 1 mas) at different elevations to estimate an 
atmospheric zenith delay error in the VLBA calibrator model (see 
\citeNP{ReidBrunthaler2004} and \citeNP{BrunthalerReidFalcke2005b} for a 
discussion). In the second and third epoch we used two blocks of these geodetic
observations before and after the phase-referencing observations. From the 
fourth epoch on, we included a third geodetic block in the middle of the 
observation.

The data were edited and calibrated using standard techniques in the 
Astronomical Image Processing System (AIPS). A-priori amplitude calibration
was applied using system temperature measurements and standard gain curves. 
Zenith delay corrections were performed based on the results of the 
geodetic-like observations. Data from the St. Croix station were flagged 
due to high phase noise in all observations. The maser in IC\,10 and 
NVSS~J002108+591132 were imaged in AIPS. All detected maser features and 
NVSS~J002108+591132 were unresolved and fit by single elliptical Gaussian 
components.

\section{Results}

\subsection{Spatial Structure}

\begin{figure}
\resizebox{\hsize}{!}{\includegraphics[angle=0]{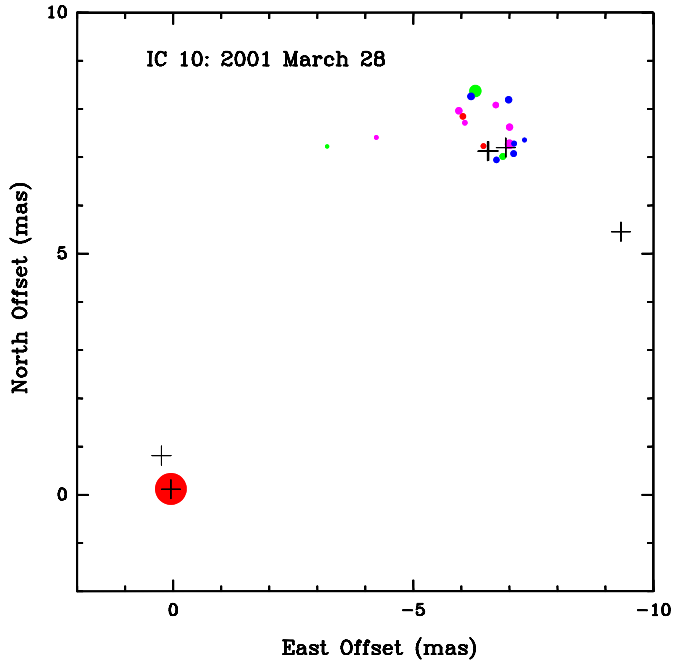}}
\caption{Composite map of the H$_2$O masers in IC\,10 from 2001 March 28. The 
area of each circle is proportional to the flux density of the respective
component. The colors denote different LSR radial velocities with 
$>-327$ km~s$^{-1}$ (red), $-327$ to $-330$ km~s$^{-1}$ (magenta), $-330$ to 
$-338$  km~s$^{-1}$ (green), and $<-338$ km~s$^{-1}$ (blue). The crosses mark 
the positions of the maser emission detected by 
\protect\citeN{ArgonGreenhillMoran1994}. The positions were aligned on the 
strongest maser component in the south-east. 
}
\label{ic10_d}
\end{figure}

In the first epoch, maser emission could be detected in 21 channels spread
over $\approx 23$~km~s$^{-1}$. The spatial distribution of the masers 
on 2001 March 28 can be seen in Fig.~\ref{ic10_d}. It is
similar to the distribution in earlier VLBI observations of IC\,10 by
\citeN{ArgonGreenhillMoran1994}. The strongest component at a LSR velocity 
of $\approx -324$~km~s$^{-1}$ is separated by $\approx10$~mas or (projected) 
6600 AU from the weaker components. This suggests that the emission is 
associated with a single object if the maser emission is similar to 
H$_2$O maser emission in Galactic star forming regions like W3(OH), W49 or 
Sgr\,B2 (e.g.~\citeNP{ReidArgonMasson1995}, ~\citeNP{WalkerBurkeJohnston1977}
and \citeNP{KobayashiIshiguroChikada1989} respectively).   The weaker
maser components form an apparent ring-like structure with a projected
size of $\approx~1.6$~mas or 1060 AU.

\subsection{Variability}

The correlated flux density of the strong feature at --324~km~s$^{-1}$ LSR 
velocity increased from 1.0 to 1.5 Jy between the first two VLBA observations
of the first epoch.
The weaker components are also very variable and they can appear or disappear 
between observations. In the observations of the second epoch, the flux 
density of the strongest component was $\sim$ 1.1 Jy and some of the weaker
components disappeared. In the third epoch, we detected only the strong 
component with a flux density of $\sim$0.7 Jy while the weak ring-like 
structure was not detected anymore. In the later epochs, the flux density 
of the remaining component dropped to $\sim$ 0.2 Jy, 0.12 Jy, and 0.07 Jy 
in the fourth, fifth, and sixth epoch, respectively. 


\subsection{Observed Motions}

\begin{figure}
\resizebox{\hsize}{!}{\includegraphics[angle=-90]{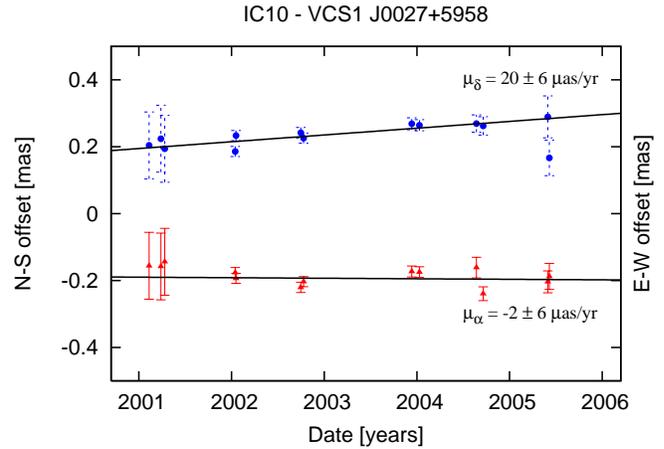}}
\caption{The position of the maser in IC\,10 relative to the phase-reference 
source VCS1 J0027+5958 in East-West (red triangles) and North-South 
(blue circles). The lines show a variance weighted linear fit to the data 
points.}
\label{pos_ic10}
\end{figure}

The position offsets 
of the strongest maser feature in IC\,10 are shown in Fig.~\ref{pos_ic10}. 
The uncertainties in the observations of the first epoch are 
larger than the others, because no geodetic-like observations were 
done to compensate the zenith delay errors. A rectilinear motion was fit to 
the data and yielded a value of --2$\pm$6 
$\mu$as~yr$^{-1}$ toward the East and 20$\pm$6 $\mu$as~yr$^{-1}$ toward the
North. 

\begin{figure}
\resizebox{\hsize}{!}{\includegraphics[angle=-90]{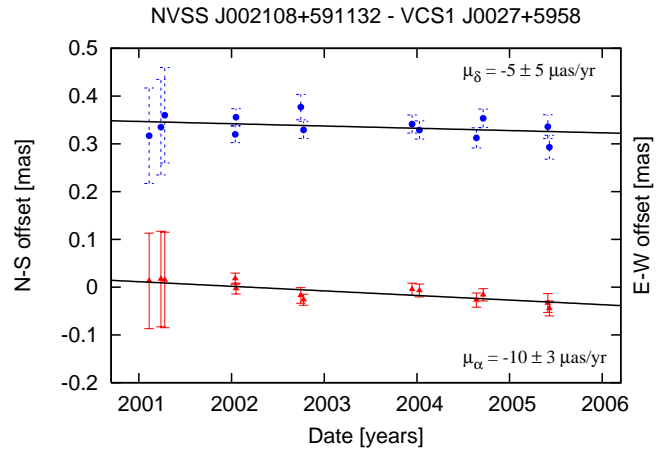}}
\caption{The position of NVSS J002108+591132 relative to the phase-reference 
source VCS1 J0027+5958 in East-West (red triangles) and North-South 
(blue circles). The lines show a variance weighted linear fit to the data 
points.}
\label{pos_a3}
\end{figure}

The position offsets of NVSS J002108+591132 are shown in Fig.~\ref{pos_a3}. 
A rectilinear motion was fit to the data and yielded a motion of 
--10$\pm$3 $\mu$as~yr$^{-1}$ toward the East and --5$\pm$5 $\mu$as~yr$^{-1}$ 
toward the North. Hence, NVSS J002108+591132 shows a small but potentially 
significant motion in right ascension. The apparent motion of NVSS 
J002108+591132 could be caused by unknown systematic errors.

The phase calibrator VCS1 J0027+5958 may have unresolved structure, e.g. 
a core-jet structure. The observed image of the source is the convolution 
of the source structure and the synthesized beam of the VLBA. Flux density 
variations of the individual components could move the position of the 
observed image by a fraction of the beam size. Since the phase calibrator 
is assumed to be stationary, this would shift the positions of all target 
sources by the same amount. 

The observed motion of NVSS J002108+591132 could also be caused by 
some errors in the geometry of the correlator model (i.e. antenna positions, 
earth orientation parameters). These errors would be 
similar for closely spaced observations, but different for observations 
separated by several months. The angular separation between IC\,10 and 
NVSS J002108+591132 (8') is much smaller than the separation between IC\,10 
and VCS1 J0027+5958 (1$^\circ$), and the position shift induced by geometric 
errors would be similar for IC\,10 and NVSS J002108+591132. 

\begin{figure}
\resizebox{\hsize}{!}{\includegraphics[angle=-90]{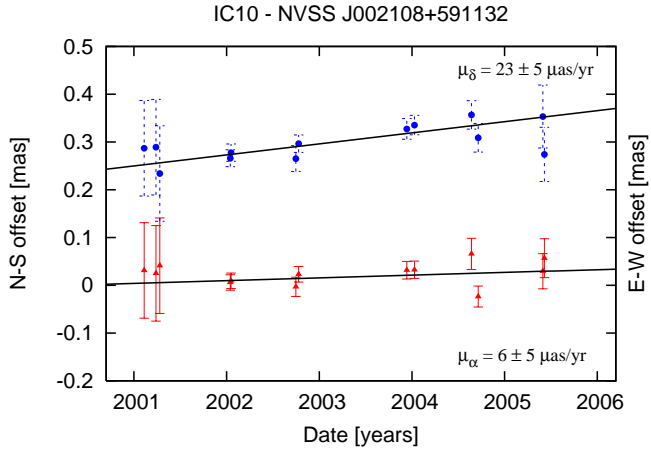}}
\caption{The position of the strongest maser feature in IC\,10 relative to NVSS J002108+591132 in East-West (red triangles) and North-South (blue circles). The lines show a variance weighted linear fit to the data points.}
\label{pos_ic10-a3}
\end{figure}

In both cases, the motion of the maser in IC\,10 relative to 
NVSS~J002108+591132 would be a better estimate of the proper motion of 
IC\,10. However, it cannot be ruled out that the apparent motion of 
NVSS~J002108+591132 is caused by an unresolved core-jet structure in 
NVSS~J002108+591132 itself. Since strong sources are expected to show more 
jet-structure than weak sources, amplitude variations of VCS1 J0027+5958 
(70--290 mJy) are larger than those in NVSS~J002108+591132 (6--11 mJy),
and the angular separation between IC\,10 and 
NVSS~J002108+591132 is much smaller than the separation between IC\,10 and 
VCS1 J0027+5958, we consider NVSS~J002108+591132 as the better astrometric 
reference source. Fig.~\ref{pos_ic10-a3} shows the position of the strongest 
maser component in IC\,10 relative to NVSS~J002108+591132. A rectilinear 
motion was fit to the data and yielded a motion of 6$\pm$5 $\mu$as~yr$^{-1}$ 
toward the East and 23$\pm$5 $\mu$as~yr$^{-1}$ toward the North and we 
will adopt these values for the proper motion of the maser.

\section{Discussion}
\subsection{Space Motion of IC\,10}

The measured proper motion $\tilde{\vec v}_{prop}$ of the maser in IC\,10 can 
be decomposed into a sum of several components, relative to a frame at rest at
the center of the Milky Way:

\begin{eqnarray}
\tilde{\vec v}_{prop}=  \vec v_{rot} + \vec v_{pec} 
+ \vec v_\odot + \vec v_{IC\,10}
\end{eqnarray}

Here $\vec v_{rot}$ is the motion of the masers due to the internal
galactic rotation in IC\,10, $\vec v_{pec}$ is the peculiar 
motion of the masers relative to circular galactic rotation and  $\vec
v_\odot$ is the apparent motion of IC\,10 caused by the rotation of the
Sun about the Galactic Center. The last contribution $\vec v_{IC\,10}$ 
is the true proper motion of the galaxy IC\,10.

The H$_2$O masers in IC\,10 are located within a massive
H\,I cloud in the central disk. If one assumes that the masers are
rotating with the disk, one can calculate its expected proper motion.
\citeN{ShostakSkillman1989} measure an inclination of 45$^\circ$ from
the ellipticity of its H\,I distribution. The masers are 33 arcseconds
(106 pc) east and 99 arcseconds (317 pc) south of the kinematic
center. Unfortunately no position angle of the major axis was
given. \citeN{WilcotsMiller1998} used higher resolution VLA
observations of the H\,I content of IC\,10 to fit a tilted ring model to
the velocity field of the disk of IC\,10. This model has a separate
rotation speed, inclination and position angle for each
ring.  They find a highly inclined disk in the inner 110 arcseconds with a
position angle of $\approx 75 ^\circ$ and a rotational velocity of
$\approx 30 $ km~s$^{-1}$. The position of the kinematic center of
their tilted ring model was not given. If one combines the kinematic center
of ~\citeN{ShostakSkillman1989} with the inclination and position angle
of ~\citeN{WilcotsMiller1998}, one gets an expected transverse motion 
for the maser ($\vec v_{rot}$) of 26 and 11 km~s$^{-1}$ toward the East and 
North, respectively. 

If one calculates the expected motion for different realistic
scenarios (i.e. changing the kinematic center by $\pm$20 arcseconds, and the
inclination and the position angle of the major axis by $\pm$20$^\circ$), one 
gets always transverse motions between 20-30 km~s$^{-1}$ toward the East and 
5-15 km~s$^{-1}$ toward the North. The deviation of the motion of the masers
from the galactic rotation is unknown. The radial velocity of the CO gas at 
the position of the maser in IC\,10 is about $-330$ km~s$^{-1}$ 
\cite{Becker1990}, which is close to the radial
velocity of the maser. In our Galaxy peculiar motions of
star forming regions can be 20 km~s$^{-1}$ as seen in W3(OH)
(\citeNP{XuReidZheng2006,HachisukaBrunthalerMenten2006}).
Hence, to be conservative, we adopt values of $25\pm20$ and $10\pm20$ 
km~s$^{-1}$ toward the East and North, respectively. This 
translates to $\dot\alpha_{rot}=8\pm6$ and $\dot\delta_{rot}=3\pm6$ 
$\mu$as~yr$^{-1}$ at a distance of 660$\pm$60~kpc.

The rotation of the Sun about the Galactic Center causes an apparent
motion of IC\,10. The motion of the Sun can be decomposed into a
circular motion of the local standard of rest (LSR) and the peculiar
motion of the Sun. The peculiar motion of the Sun has been determined
from Hipparcos data by ~\citeN{DehnenBinney1998} to be in km~s$^{-1}$
U$_0$=10.00$\pm$0.36 (radially inwards), V$_0$=5.25$\pm$0.62 (in the
direction of Galactic rotation) and W$_0$=7.17$\pm$0.38 (vertically
upwards). VLBI measurements of the proper motion of SgrA*, the compact 
radio source at the Galactic Center, yield a motion of 6.379 $\pm$ 0.026 
mas~yr$^{-1}$ along the Galactic plane 
\cite{ReidReadheadVermeulen1999,ReidBrunthaler2004}. Combined with a recent 
geometric distance estimate of the Galactic Center of 7.62 $\pm$ 0.32 kpc 
\cite{EisenhauerGenzelAlexander2005}, one gets a circular velocity of 
225$\pm$10 km~s$^{-1}$ for the LSR.

This motion of the Sun causes an apparent
proper motion of $38\pm4~\mu$as~yr$^{-1}$ in Galactic longitude and
-6$\pm$1 $\mu$as~yr$^{-1}$ in Galactic latitude  
(for a distance of 660 kpc and Galactic
coordinates of IC\,10 of $l=118.96^\circ$, $b=-3.32^\circ$). Converted to
equatorial coordinates, one gets
$\dot\alpha_{\odot}=37\pm4~\mu$as~yr$^{-1}$  and
$\dot\delta_{\odot}=-11\pm1~\mu$as~yr$^{-1}$.

The true proper motion of IC\,10 is then given by

\begin{eqnarray}
\nonumber\dot\alpha_{IC\,10}&=&\dot{\tilde\alpha}_{prop} - \dot\alpha_{rot} -\dot\alpha_\odot\\\nonumber
&=&(6~(\pm5)-8~(\pm6)-37~(\pm4))~\mu\mathrm{as}~\mathrm{yr}^{-1}\\\nonumber
&=&-39\pm9~\mu\mathrm{as}~\mathrm{yr}^{-1}
=-122\pm31~\mathrm{km}~\mathrm{s}^{-1}\\
\mathrm{and}\\\nonumber
\dot\delta_{IC\,10}&=&\dot{\tilde\delta}_{prop} - \dot\delta_{rot} -\dot\delta_\odot\\\nonumber
&=& (23~(\pm5)-3~(\pm6)+11~(\pm1))~\mu\mathrm{as}~\mathrm{yr}^{-1}\\\nonumber
&=&31\pm8~\mu\mathrm{as}~\mathrm{yr}^{-1}=97\pm27~{\mathrm{km}}~\mathrm{s}^{-1}
\end{eqnarray}

The measured systematic heliocentric velocity of IC\,10 ($-344\pm3$
km~s$^{-1}$,~\citeNP{deVaucouleursdeVaucouleursCorwins1991}) is the sum
of the radial motion of IC\,10 toward the Sun and the component of the
solar motion about the Galactic Center toward IC\,10 which is
-196$\pm$10 km~s$^{-1}$. Hence IC\,10 is moving with 148$\pm$10
km~s$^{-1}$ toward the Sun.

\begin{figure}
\resizebox{0.7\hsize}{!}{\includegraphics[bbllx=2.5cm,bburx=17.5cm,bblly=10.5cm,bbury=27cm,clip=,angle=0]{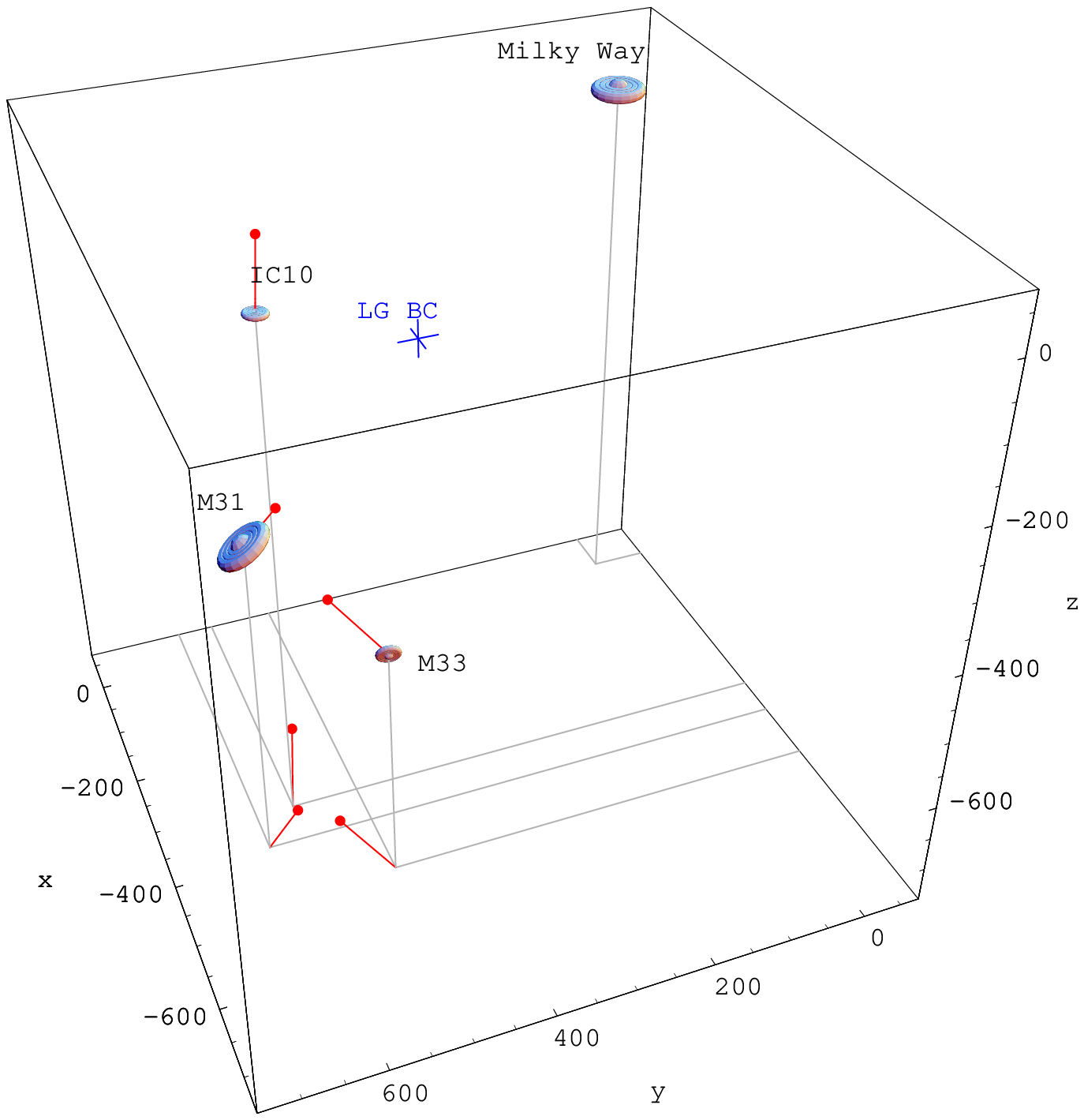}}
\resizebox{0.7\hsize}{!}{\includegraphics[bbllx=2.5cm,bburx=17.5cm,bblly=10.5cm,bbury=27cm,clip=,angle=0]{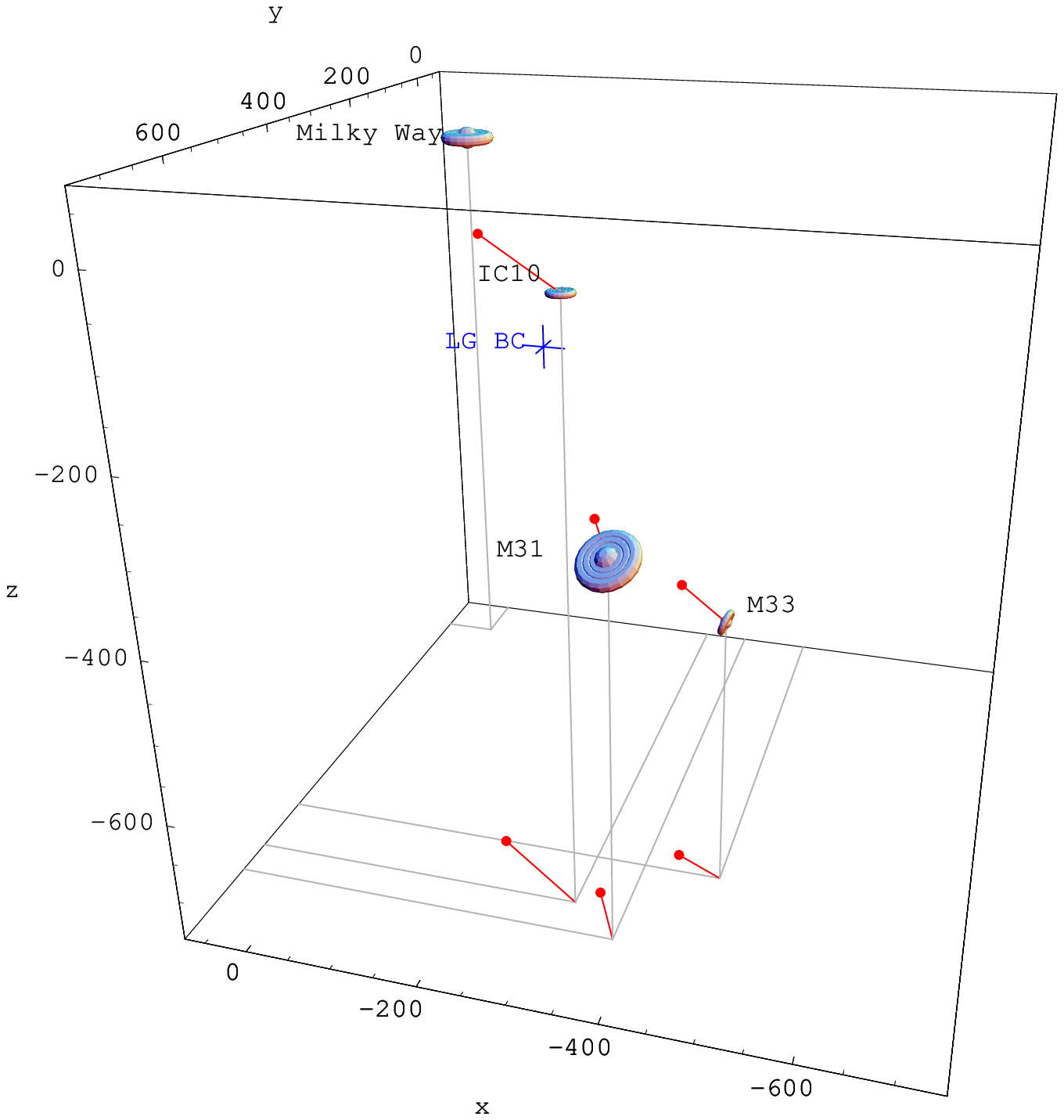}}
\caption{Schematic view of the Local Group from two viewing angles with the 
space velocity of IC\,10 and M33 (the latter taken from 
\protect\citeNP{BrunthalerReidFalcke2005}) and the radial velocity of 
Andromeda relative to the Milky Way. The blue cross marks the position 
of the Local Group Barycenter (LG BC) according to  
\protect\citeN{vandenBergh1999}.}
\label{LG}
\end{figure}

The proper motion and the radial velocity combined give the
three-dimensional space velocity of IC\,10. The total velocity is
215$\pm$42~km~s$^{-1}$ relative to the Milky Way. This velocity vector is 
shown in the schematic view of the Local Group in Fig.~\ref{LG}. Here, we 
used Cartesian coordinates, where the Sun is located at the origin and the
Galactic Center is located at (x,y,z)=(7.62,0,0) (see Appendix~\ref{trans}
for details).

\subsection{Local Group Dynamics and Mass of M31}

If IC\,10 or M33 are bound to M31, then the velocity of the two galaxies 
relative to M31 must be smaller than the escape velocity and one can 
deduce a lower limit on the mass of M31:

\begin{eqnarray}
M_{M31}>\frac{v_{rel}^2R}{2G}.
\end{eqnarray}

\begin{figure}
\center{M$_\mathrm{M31}$ [M$_\odot$]}
\resizebox{0.9\hsize}{!}{\includegraphics[bbllx=0cm,bburx=35.5cm,bblly=1.0cm,bbury=5.5cm,clip=,angle=0]{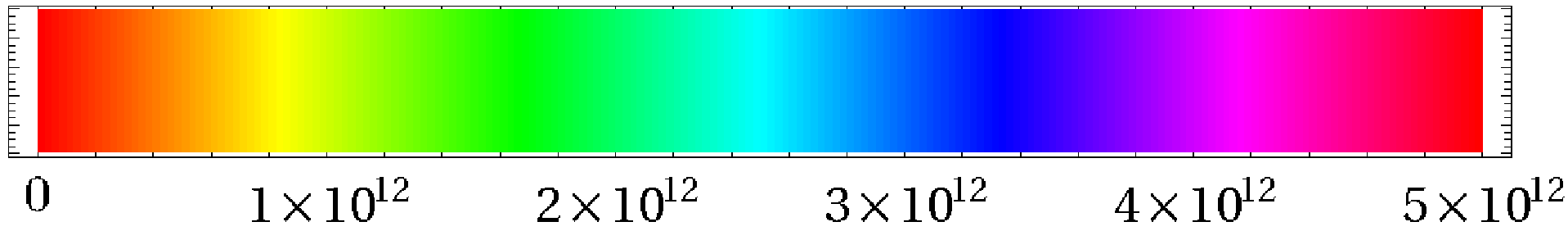}}
\resizebox{0.8\hsize}{!}{\includegraphics[bbllx=2.5cm,bburx=13cm,bblly=16.9cm,bbury=25.5cm,clip=,angle=0]{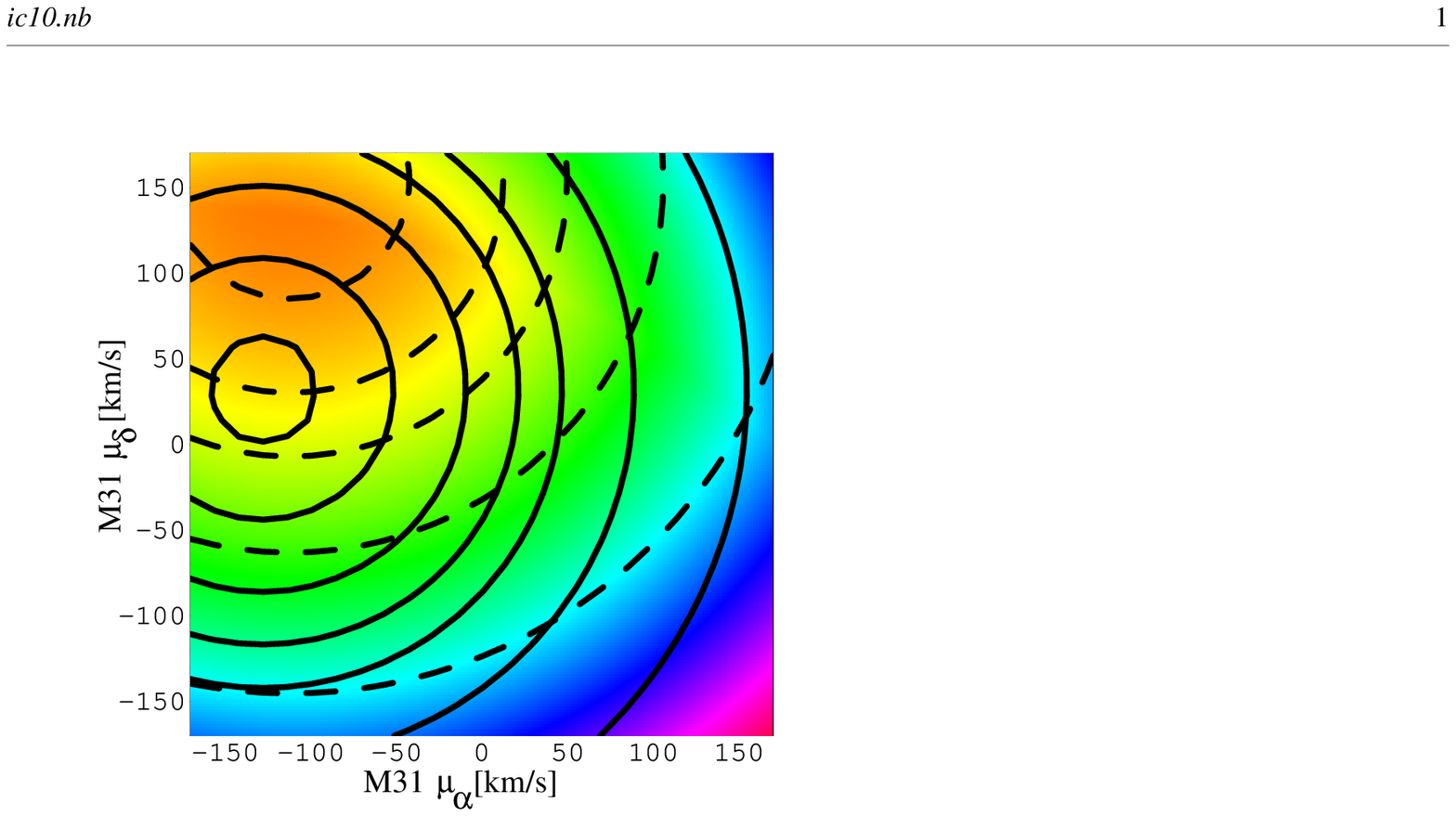}}
\resizebox{0.8\hsize}{!}{\includegraphics[bbllx=2.5cm,bburx=13cm,bblly=16.9cm,bbury=25.5cm,clip=,angle=0]{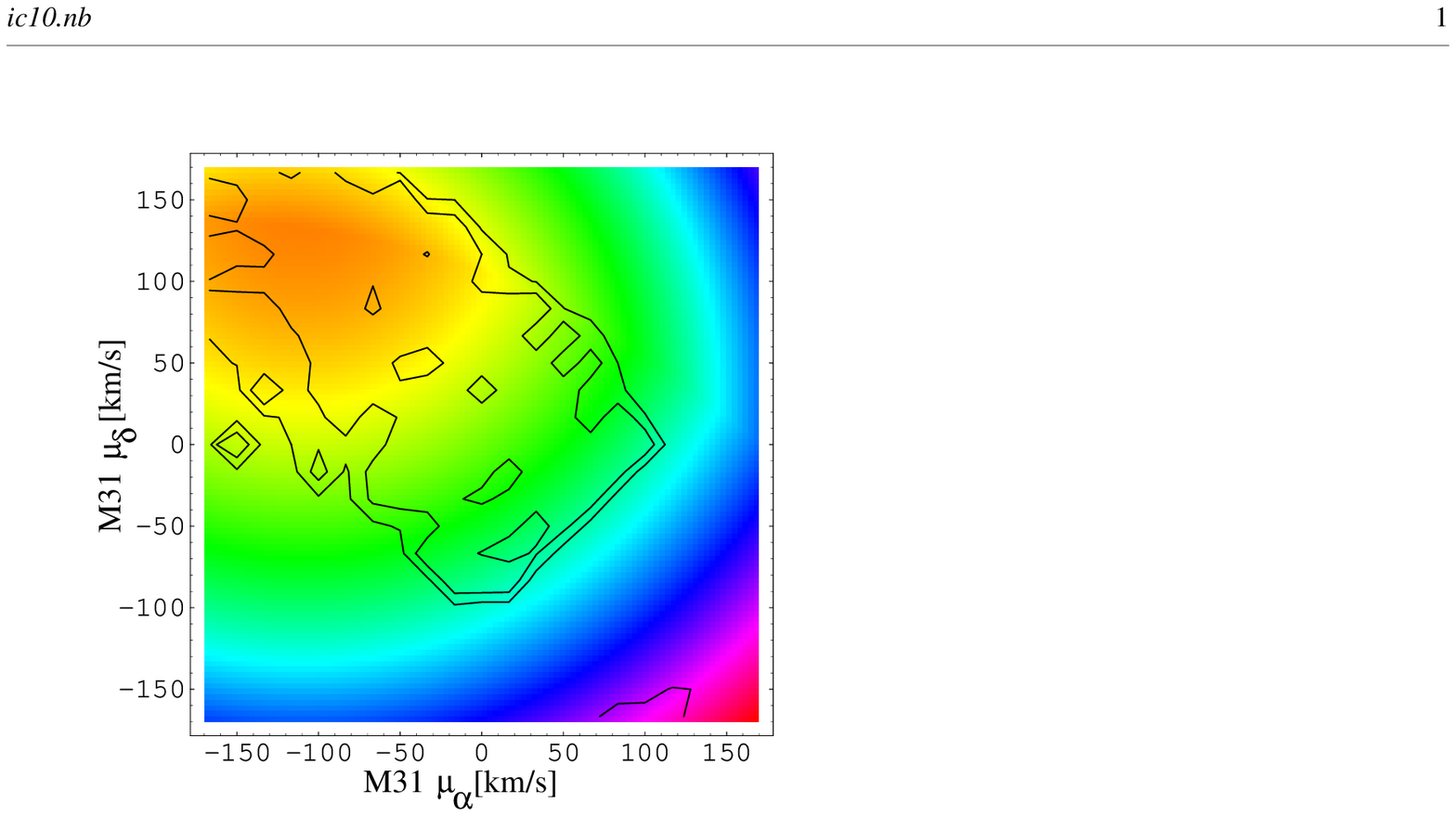}}
\caption{{\bf Top:} Lower limit on the mass of M31 for different tangential motions of M31
assuming that M33 (dashed) or IC\,10 (solid) are bound to M31. The lower limits
are (4, 5, 7.5, 10, 15, 25)$\times10^{11}$M$_\odot$ for M33, and (0.7, 1, 2.5, 
5, 7.5, 10, 15, 25)$\times10^{11}$M$_\odot$ for IC\,10, rising from inside. 
The colour scale indicates the maximum of both values.
{\bf Bottom:} The colour scale is the same as above and gives the 
lower limit on the mass of M31. The contours show ranges of proper
motions that would have lead to a large ammount of stars stripped from the disk
of M33 through interactions with M31 or the Milky Way in the past.
The contours delineate 20\% and 50\%  of the total number of stars stripped 
\protect\cite{LoebReidBrunthaler2005}. These regions can be excluded, since the
stellar disk of M33 shows no signs of such interactions.}
\label{mass-m31}
\end{figure}

A relative velocity of 147 km~s$^{-1}$ -- for a zero tangential motion of M31 
-- and a distance of 262 kpc between IC\,10 and M31, gives a lower limit of 
6.6 $\times 10^{11}$M$_\odot$. One can repeat this calculation for any 
tangential motion of M31. The results are shown in Fig.~\ref{mass-m31} (top). 
The lowest 
value of 0.7 $\times 10^{11}$M$_\odot$ is found for a tangential motion of M31 
of --130 km~s$^{-1}$ toward the East and 35 km~s$^{-1}$ toward the North.


For a relative motion of 230 km~s$^{-1}$ between M33 and M31 -- again for a 
zero tangential motion of M31 -- and a distance of 202 kpc, one gets a lower 
limit of 1.2 $\times 10^{12}$M$_\odot$ \cite{BrunthalerReidFalcke2005}. 
Fig.~\ref{mass-m31} (top) shows also the lower limit of the mass of M31 for 
different 
tangential motions of M31 if M33 is bound to M31. The lowest value is 
4 $\times 10^{11}$M$_\odot$ for a tangential motion of M31 of 
--115 km~s$^{-1}$ toward the East and 160 km~s$^{-1}$ toward the North.

\citeN{LoebReidBrunthaler2005} 
find that proper motions of M31 in negative right ascension and positive 
declination would have lead to close interactions between M31 and M33 in the
past. These proper motions of M31 can be ruled out, since the stellar disk of 
M33 does not show any signs of strong interactions. 
\citeN{LoebReidBrunthaler2005} used a total mass of M31 of 
3.4$\times10^{12}$M$_\odot$ in their simulations. Although simulations with 
lower masses of M31 yield weaker interactions, motions in negative right 
ascension and positive declination are still ruled out.

Thus, we can rule out these
regions in Fig.~\ref{mass-m31}. This results in a lower limit of 
7.5$\times 10^{11}$M$_\odot$ for M31 and agrees with a recent
estimate of $12.3^{+18}_{-6}\times10^{11}$~M$_\odot$ derived from the
three-dimensional positions and radial velocities of its satellite
galaxies \cite{EvansWilkinson2000}.


\section{Summary}
We have presented astrometric VLBA observations of the H$_2$O maser in the
Local Group galaxy IC\,10. We detected a ring-like structure in one epoch with 
a projected diameter of $\sim$1060 AU. We measured the proper motion of the 
maser relative to two background quasars. Correcting for the internal rotation 
of IC\,10 and the rotation of the Milky Way this measurement yields a proper 
motion of --39$\pm$9 $\mu$as~yr$^{-1}$ toward the East and 31$\pm$8 
$\mu$ as~yr$^{-1}$ toward the North, which corresponds to a total space 
velocity of 215$\pm$42 km~s$^{-1}$ for IC\,10 relative to the 
Milky Way. If IC\,10 and M33 are bound to M31, one can calculate a lower limit 
of the mass of M31 of 7.5~$\times 10^{11}$M$_\odot$.

\begin{acknowledgements}
This research was supported by the DFG Priority Programme 1177.
\end{acknowledgements}

\bibliography{brunthal_refs}
\bibliographystyle{aa}

\appendix{}

\section{Coordinate Transformations}
\label{trans}

We define a Cartesian coordinate system where the Sun is located at the origin.
To convert from Galactic to Cartesian coordinates, we used

\begin{eqnarray}
\left(\begin{array}{lcr}x\\y\\z\end{array}\right)=
\left(\begin{array}{lcr}R~\mathrm{cos}~l~\mathrm{cos}~b\\R~\mathrm{sin}~l~\mathrm{cos}~b\\R~\mathrm{sin}~b\end{array}\right),
\end{eqnarray}

where $R$ is the distance, $l$ is the Galactic Longitude, and $b$ is the 
Galactic Latitude.

To convert the proper motions in right ascension ($\dot\alpha$) and 
declination ($\dot\delta$) into proper motions in Galactic Longitude ($\dot l$)
and Galactic Latitude ($\dot b$), we used

\begin{eqnarray}
\left(\begin{array}{lcr}\dot l\\\dot b\end{array}\right)=\left(\begin{array}{lcr}\mathrm{cos}~\theta&\mathrm{sin}~\theta\\-\mathrm{sin}~\theta&\mathrm{cos}~\theta\end{array}\right)\left(\begin{array}{lcr}\dot \alpha\\\dot\delta\end{array}\right),
\end{eqnarray}

where $\theta$ is the required rotation angle at the position of the source 
(i.e. $\theta_{IC\,10}=-6.94^\circ$, $\theta_{M33}=-348.95^\circ$, 
$\theta_{M31}=-2.74^\circ$).

To compute the velocities in Cartesian coordinates, we constructed for
each source three orthogonal vectors $\vec e_r$, $\vec e_l$, and $\vec e_b$ 
given by

\begin{eqnarray}
\vec e_r=\left(\begin{array}{lcr}x\\y\\z\end{array}\right);~~
\vec e_l=\left(\begin{array}{c}\frac{y}{x}\\-1\\0\end{array}\right);~~
\vec e_b=\vec e_r\times \vec e_l.
\end{eqnarray}

These vectors can be normalized to give three orthogonal unit vectors 
$\hat{\vec e_r}$, $\hat{\vec e_l}$, and $\hat{\vec e_b}$. Then the total 
velocity $\vec v_\mathrm{tot}$ of an object with proper motions $\dot l$, 
$\dot b$, and the radial velocity $v_\mathrm{rad}$ is

\begin{eqnarray}
\vec v_\mathrm{tot}=\dot l~\hat{\vec e_l} + \dot b~\hat{\vec e_b} 
+ v_\mathrm{rad} \hat{\vec e_r}.
\end{eqnarray}

\end{document}